\documentclass[a4paper]{spie}  

\newcommand{\Hitomi}{\textit{Hitomi}}

\usepackage{amsmath,amsfonts,amssymb}
\usepackage[dvipdfmx]{graphicx}
\usepackage[colorlinks=true, allcolors=blue]{hyperref}
\usepackage{aas_macros}
\usepackage{threeparttable}

\title{Status of Xtend telescope onboard X-Ray Imaging and Spectroscopy Mission (XRISM)}

\author[a]{Koji Mori}
\affil[a]{Faculty of Engineering, University of Miyazaki, 1-1 Gakuen Kibanadai Nishi, Miyazaki, Miyazaki 889-2192, Japan}
\author[b]{Hiroshi Tomida}
\affil[b]{Japan Aerospace Exploration Agency, Institute of Space and Astronautical Science, 3-1-1 Yoshino-dai, Chuo-ku, Sagamihara, Kanagawa 252-5210, Japan}

\author[c]{Hiroshi Nakajima}
\affil[c]{College of Science and Engineering, Kanto Gakuinn University, Kanazawa-ku, Yokohama, Kanagawa 236-8501, Japan}

\author[d]{Takashi Okajima}
\affil[d]{NASA's Goddard Space Flight Center, Greenbelt, MD 20771, USA}

\author[e]{Hirofumi Noda}
\affil[e]{Astronomical Institute, Tohoku University, 6-3 Aramakiazaaoba, Aoba-ku, Sendai, Miyagi 980-8578, Japan}

\author[f]{Hiroyuki Uchida}
\affil[f]{Department of Physics, Kyoto University, Kitashirakawa Oiwake-cho,Sakyo-ku, Kyoto, Kyoto 606-8502, Japan}

\author[b]{Hiromasa Suzuki}

\author[g]{Shogo Benjamin Kobayashi}
\affil[g]{Department of Physics, Faculty of Science, Tokyo University of Science, Kagurazaka, Shinjuku-ku, Tokyo 162-0815, Japan}

\author[h]{Tomokage Yoneyama}
\affil[h]{Faculty of Science and Engineering, Chuo University, 1-13-27 Kasuga, Bunkyo, Tokyo 112-8551, Japan}

\author[i]{Kouichi Hagino}
\affil[i]{Department of Physics, University of Tokyo, 7-3-1 Hongo, Bunkyo, Tokyo 113-0033, Japan}

\author[j]{Kumiko Nobukawa}
\affil[j]{Department of Physics, Kindai University, 3-4-1 Kowakae, Higashi-Osaka, Osaka 577-8502, Japan}

\author[k]{Takaaki Tanaka}
\affil[k]{Department of Physics, Konan University, 8-9-1 Okamoto, Higashinada, Kobe,
Hyogo 658-8501}

\author[l]{Hiroshi Murakami}
\affil[l]{Faculty of Informatics, Tohoku Gakuin University, 3-1 Shimizukoji, Wakabayashi-ku, Sendai, Miyagi 984-8588}

\author[m]{Hideki Uchiyama}
\affil[m]{Science Education, Faculty of Education, Shizuoka University, Suruga-ku, Shizuoka, Shizuoka 422-8529, Japan}

\author[n]{Masayoshi Nobukawa}
\affil[n]{Faculty of Education, Nara University of Education, Nara, Nara 630-8528, Japan}

\author[o,p]{Hironori Matsumoto}
\affil[o]{Department of Earth and Space Science, Osaka University, 1-1 Machikaneyama-cho, Toyonaka, Osaka 560-0043, Japan}
\affil[p]{Forefront Research Center, Osaka University, 1-1 Machikaneyama-cho, Toyonaka, Osaka 560-0043, Japan}

\author[f]{Takeshi Tsuru}

\author[a]{Makoto Yamauchi}
\author[a]{Isamu Hatsukade}

\author[o,p]{Hirokazu Odaka}

\author[q]{Takayoshi Kohmura}
\affil[q]{Department of Physics, Faculty of Science and Technology, Tokyo University of Science, 2641 Yamazaki, Noda, Chiba 270-8510, Japan}

\author[r]{Kazutaka Yamaoka}
\affil[r]{Department of Physics, Nagoya University, Chikusa-ku, Nagoya,	Aichi 464-8602, Japan}

\author[b]{Manabu Ishida}
\author[b]{Yoshitomo Maeda}
\author[f,s]{Takayuki Hayashi}
\author[f,s]{Keisuke Tamura} 
\author[f,s]{Rozenn Boissay-Malaquin} 
\affil[s]{Center for Space Science and Technology, University of Maryland, Baltimore County (UMBC), Baltimore, MD 21250, USA}
\author[t]{Toshiki Sato}
\affil[t]{Department of Physics, School of Science and Technology, Meiji University, 1-1-1 Higashi Mita, Tama-ku, Kawasaki, Kanagawa 214-8571, Japan}

\author[u]{Tessei Yoshida}
\affil[u]{Japan Aerospace Exploration Agency, Institute of Space and Astronautical Science, 2-1-1, Sengen, Tsukuba, Ibaraki 305-8505, Japan}

\author[b]{Yoshiaki Kanemaru}
\author[v]{Junko Hiraga}
\affil[v]{Department of Physics, Kwansei Gakuin University, 2-2 Gakuen, Sanda, Hyogo 669-1337, Japan}
\author[b,w]{Tadayasu Dotani}
\affil[w]{Department of Space and Astronautical Science, School of Physical Sciences,
	SOKENDAI (The Graduate University for Advanced Studies),
	3-1-1 Yoshino-dai, Chuou-Ku, Sagamihara, Kanagawa 252-5210, Japan}
\author[x]{Masanobu Ozaki}
\affil[x]{Advanced Technology Center, National Astronomical Observatory of Japan, 2-21-1 Osawa, Mitaka, Tokyo 181-8588, Japan}
\author[o]{Hiroshi Tsunemi}

\author[f]{Shun Inoue}

\author[m]{Ryuishi Azuma}
\author[k]{Yuma Aoki}
\author[c]{Yoh Asahina}
\author[c]{Shotaro Nakamura}
\author[c]{Takamitsu Kamei}
\author[c]{Masahiro Fukuda}

\author[o]{Kazunori Asakura}
\author[o]{Marina Yoshimoto}
\author[o]{Yuichi Ode}
\author[o]{Tomohiro Hakamata}
\author[o]{Mio Aoyagi}
\author[o]{Kohei shima}

\author[j]{Yuma Aoki}
\author[j]{Yamato Ito}

\author[q]{Daiki Aoki}
\author[q]{Kaito Fujisawa}
\author[q]{Yasuyuki Shimizu}
\author[q]{Mayu Higuchi}

\author[a]{Keitaro Miyazaki}
\author[a]{Kohei Kusunoki}
\author[a]{Yoshinori Otsuka}
\author[a]{Haruhiko Yokosu}
\author[a]{Wakana Yonemaru}
\author[a]{Kazuhiro Ichikawa}
\author[a]{Hanako Nakano}
\author[a]{Reo takemoto}
\author[a]{Tsukasa Matsushima}

\author[o]{Kiyoshi Hayashida}

\authorinfo{Further author information: (Send correspondence to K.M.)\\K.M.: E-mail: mori@astro.miyazaki-u.ac.jp, Telephone: 81 985 58 7371}
 
\begin{document} 
\maketitle

\begin{abstract}
Xtend is one of the two telescopes onboard the X-ray imaging and spectroscopy
mission (XRISM), which was launched on September 7th, 2023. Xtend comprises the Soft
X-ray Imager (SXI), an X-ray CCD camera, and the X-ray Mirror Assembly (XMA), a
thin-foil-nested conically approximated Wolter-I optics. A large field of view of
$38^{\prime}\times38^{\prime}$ over the energy range from 0.4 to 13~keV is realized
by the combination of the SXI and XMA with a focal length of 5.6~m. The SXI employs
four P-channel, back-illuminated type CCDs with a thick depletion layer of
200~$\mu$m. The four CCD chips are arranged in a 2$\times$2 grid and cooled down to
$-110$ $^{\circ}$C with a single-stage Stirling cooler. Before the launch of XRISM,
we conducted a month-long spacecraft thermal vacuum test. The performance
verification of the SXI was successfully carried out in a course of multiple thermal
cycles of the spacecraft. About a month after the launch of XRISM, the SXI was
carefully activated and the soundness of its functionality was checked by a
step-by-step process. Commissioning observations followed the initial operation. We
here present pre- and post-launch results verifying the Xtend performance. All the
in-orbit performances are consistent with those measured on ground and satisfy the
mission requirement. Extensive calibration studies are ongoing.
\end{abstract}


\section{INTRODUCTION}
\label{sec:intro}  

X-Ray Imaging and Spectroscopy Mission (XRISM) was successfully launched on
September 7th, 2023 as the Japan's seventh X-ray astronomical
observatory\cite{TashiroSPIE2024}. The prime objectives of XRISM are to reveal
material circulation and energy transfer in cosmic plasmas and to elucidate cosmic
structures and objects\cite{2020SPIE11444E..22T}. XRISM pursues these objectives by
means of the high-resolution X-ray spectroscopy with imaging, which is realized by
two soft X-ray telescopes. One is a soft X-ray ``spectroscopy'' telescope, named
Resolve\cite{2018JLTP..193..991I}, and the other is a soft X-ray ``imaging''
telescope, named Xtend\cite{2018SPIE10699E..23H}. Both of them comprise an X-ray
Mirror Assembly (XMA), a thin-foil-nested conically approximated Wolter-I optics,
and a focal-plane detector with the same focal length of 5.6~m. The two XMAs share
the same design and show almost the same imaging performance\cite{HayashiSPIE2024,
TamuraSPIE2024}. Resolve is equipped with an X-ray micro-calorimeter array, a
non-dispersive instrument achieving unprecedented, 7~eV (or better) energy
resolution, and provides us with the high-resolution X-ray spectroscopy for diffuse
sources as well as point sources\cite{KellySPIE2024, PorterSPIE2024}. Xtend is
equipped with an X-ray CCD camera, named Soft X-ray Imager (SXI) and realizing a
large field of view (FoV) of $38^{\prime}\times38^{\prime}$, and provides us with
the wide-field X-ray imaging spectroscopy. Resolve and Xtend point toward the same
direction and the Xtend's FoV entirely covers the Resolve's limited FoV of
$3^{\prime}\times3^{\prime}$. The Xtend's pixel scale of $1.\!\!^{\prime\prime}77$
is finer than the Resolve's pixel scale of $30^{\prime\prime}$ and fully
over-samples the point spread function (PSF), whose half power diameter (HPD) is
about $1.\!\!^{\prime}3$\cite{HayashiSPIE2024, TamuraSPIE2024}. With these advanced
imaging capabilities, Xtend monitors possible flux contamination to Resolve spectra
from time-variable bright sources located outside the Resolve FoV, estimates local
diffuse background from Galactic diffuse X-ray emission and/or solar wind charge
exchange emission, and decomposes spatial structures within the Resolve's FoV. On
the other hand, the Resolve's high-resolution spectrum gives definitive information
to the interpretation of the Xtend's wide-field imaging spectroscopy. In one word,
Xtend extends the Resolve's potential to the maximum and Resolve resolves the
Xtend's spectrum into unprecedented details.

\begin{figure} [ht]
 \begin{minipage}{0.49\hsize}
  \vspace{0.7cm}
  \centering \includegraphics[width=\textwidth]{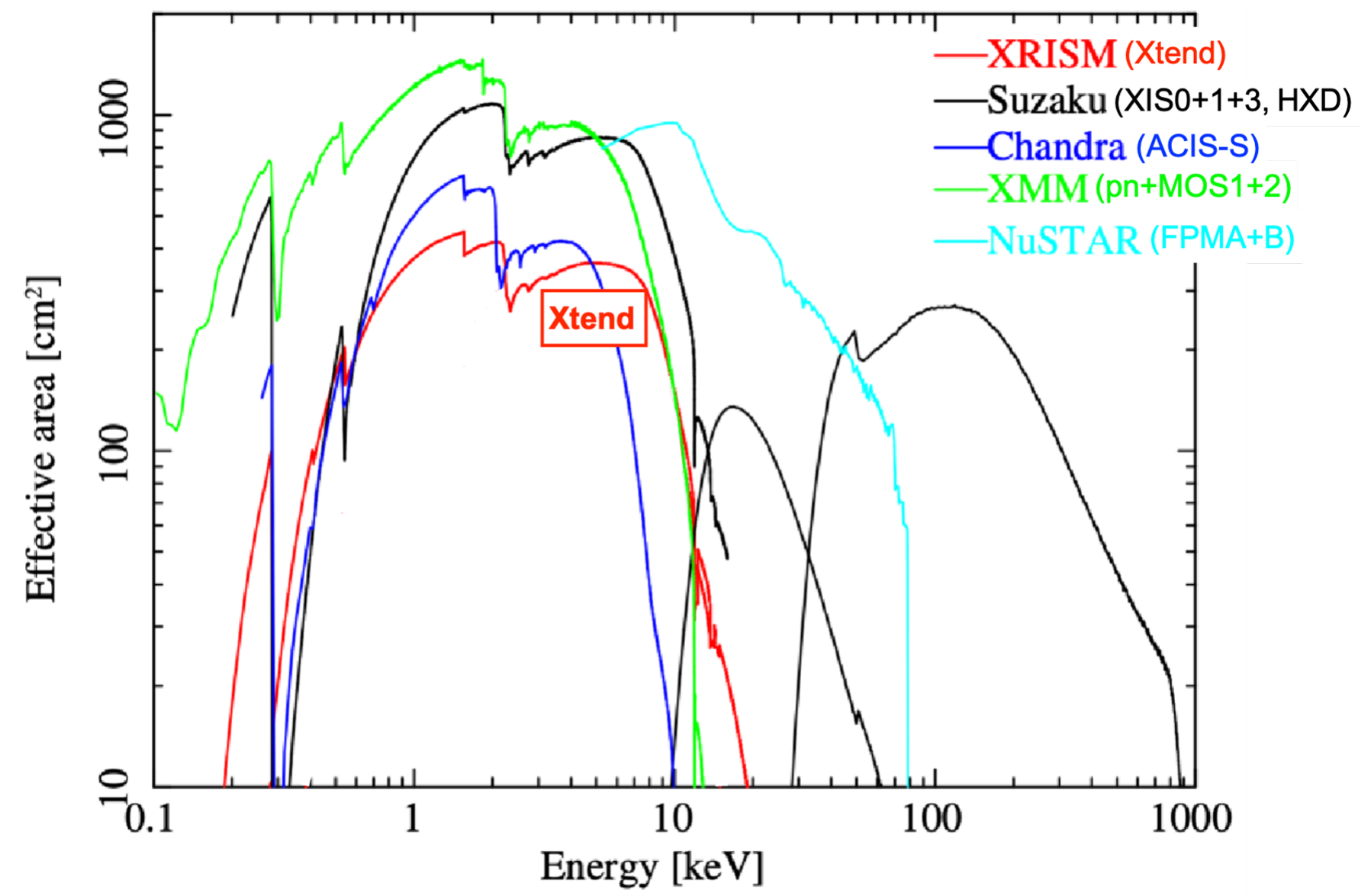}
  \vspace{0.01cm}
  \caption[]{\label{fig:EA} On-axis effective area of Xtend (red). Those of Suzaku
  (XIS0+XIS1+XIS3 and HXD; black), Chandra (ACIS-S; blue), XMM-Newton (pn+MOS1+MOS2;
  green), and NuSTAR (FPMA+FPMB; cyan) are also shown for comparison.}
 \end{minipage}
 \hspace{0.01\hsize}
  \begin{minipage}{0.49\hsize}
   \centering \includegraphics[width=\textwidth]{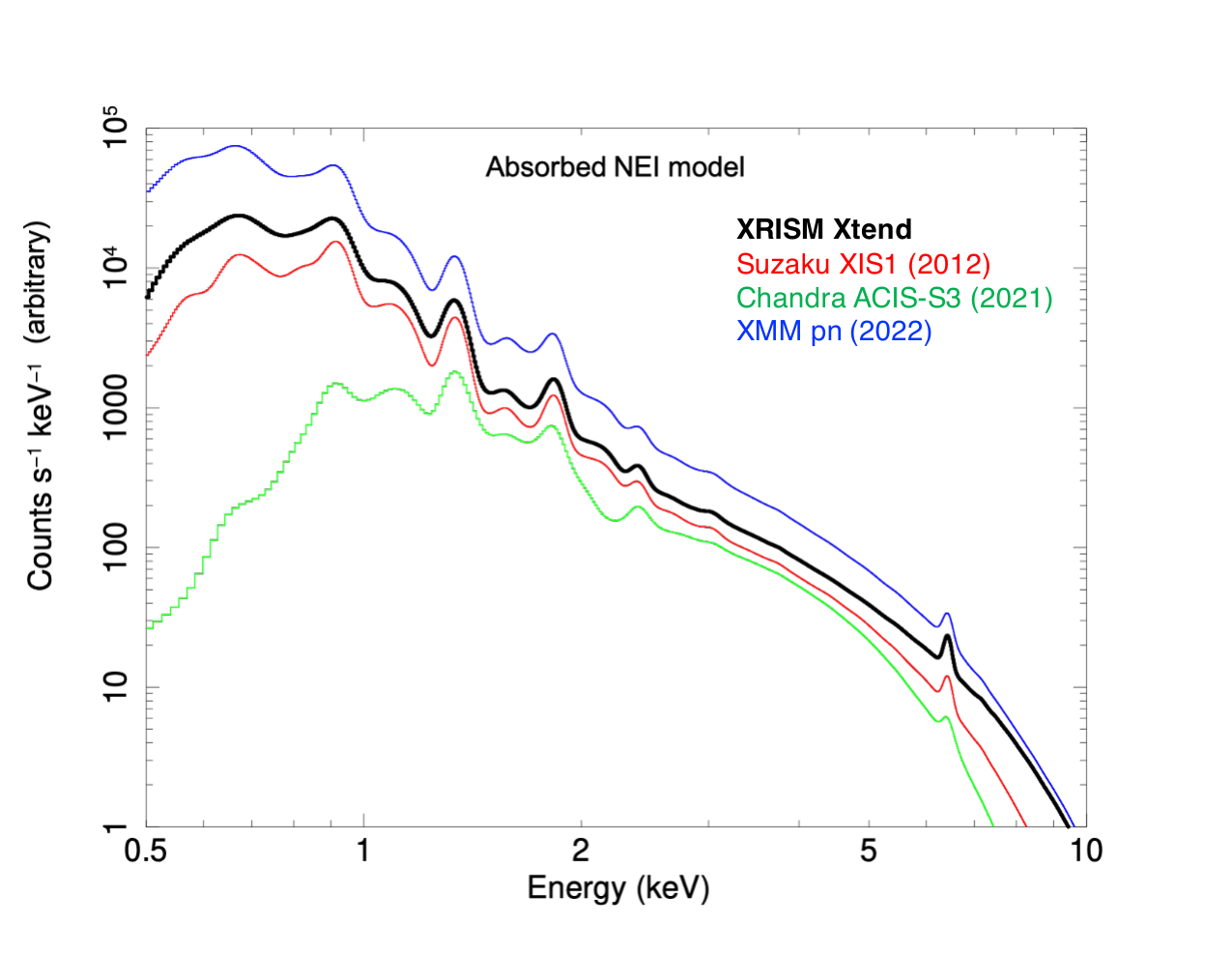}
   \caption[]{\label{fig:response} A simulated spectrum with an absorbed NEI model
   of Hitomi SXI, which is comparable to Xtend (black). Those of Suzaku XIS1 (as of
   2012; red), Chandra ACIS-S3 (as of 2021; green), and XMM-Newton PN (as of 2022;
   blue) are also shown for comparison. }
 \end{minipage}
\end{figure}

In comparison with the focal-plane CCD cameras flown in orbit so far, Xtend is
characterized with several key features. The first one is a superior X-ray response
in the high-energy band, especially in the 5-10~keV band in which X-ray emission
lines from Fe-group elements can be observed. Fig.~\ref{fig:EA}
and~\ref{fig:response} show Xtend's on-axis effective area and simulated spectrum
with an absorbed NEI model, respectively. The SXI employs P-channel back-illuminated
type CCDs with a depletion layer thickness of 200~$\mu$m\cite{ozawa_2007,
Takagi_2007, Ueda_2011}, following the \Hitomi\ CCD
camera\cite{2018JATIS...4a1211T}. Thanks to SXI's high quantum efficiency and XMA's
large collecting area, Xtend has a large effective area in the high-energy band. A
Gaussian-like simple energy response of the SXI CCD\cite{2022SPIE12181E..1TM} in
this band is also advantageous to reduce systematic uncertainties in the response
function. This superior X-ray response in the high-energy band is useful for
spatially-resolved spectroscopy of diffuse X-ray sources showing X-ray line
emissions from Fe-group elements like young supernova remnants. The second one is
its low, stable, and clean non-X-ray background (NXB), especially in the high energy
band\cite{2018PASJ...70...21N}. It is noteworthy that the background spectrum
contains no strong lines overlapping major lines from astronomical sources. These
properties of the Xtend background are mainly due to the fact that XRISM flies in a
low earth orbit, whose radiation environment is relatively low and stable, and the
material design of the CCD and detectors. This low, stable, and clean background is
useful for detection of faint diffuse X-ray emissions like those from outskirts of
galaxy clusters. The third one is the largest grasp at 7~keV among the focal-plane
CCD cameras\cite{2017NIMPA.873...16N}. Grasp is defined as the product of effective
area multiplied by effective FoV, and considered as a figure of merit to capture
diffuse X-ray emission. This feature is useful for studies of hard X-ray diffuse
sources and also transient diffuse emission like solar wind charge exchange
emission. The forth one is its large FoV, allowing serendipitous source detection in
the surrounding area of target sources. Utilizing this feature, the Xtend Transient
Search program is in operation, in which a transient source search is performed for
down-linked Xtend data once a day\cite{TsuboiSPIE2024}. All these key features,
combined with a long-exposure nature of XRISM observations, will derive new results
even from Xtend data alone.

Before the launch of XRISM, we conducted a month-long spacecraft thermal vacuum
test. The performance verification of the SXI was successfully carried out in a
course of multiple thermal cycles of the spacecraft. About a month after the launch
of XRISM, the SXI was carefully activated and the soundness of its functionality was
checked by a step-by-step process.  In this paper, we report the pre- and
post-launch status of Xtend, mostly focusing on the SXI. The calibration studies of
the XMA are presented elsewhere\cite{Boissay-MalaquinSPIE2024, TamuraSPIE2024,
HayashiSPIE2024}.  The details of the SXI initial operation are presented separately
in Ref.~\citenum{SuzukiSPIE2024}.  CCD spectra shown below are all made with grade
0, 2, 3, 4, and 6 events unless otherwise indicated. 

\section{Specifications and Observation modes}
\label{sec:spec}

\begin{center}       
 \begin{threeparttable}[ht]
  \caption{Specifications of Xtend} \label{tab:Xtend-spec}
  \begin{tabular}{|l|l|}
   \hline
   Field of view & $38^{\prime}\times38^{\prime}$ \\
   Format & 640 $\times$ 640 logical pixels per 1CCD\\
   Pixel scale & $1.\!\!^{\prime\prime}77 \times 1.\!\!^{\prime\prime}77$\\
   Energy range & 0.4--13~keV \\
   Energy resolution & $\sim$180~eV (FWHM) at 5.9~keV \\
   Effective area$^{\dagger}$ & $\sim$430~cm$^{2}$ at 1.5~keV \\
                  & $\sim$350~cm$^{2}$ at 6~keV \\
   Frame cycle & 4 seconds \\
   Charge injection & every 80 logical rows \\
   \hline 
  \end{tabular}
\begin{tablenotes}
\item[$\dagger$] to be updated by in-orbit calibration
\end{tablenotes}
 \end{threeparttable}
\end{center}

Xtend's SXI and XMA have basically identical designs to the \Hitomi\ X-ray CCD
camera\cite{2018JATIS...4a1211T} and X-ray mirror\cite{2018JATIS...4a1213I},
respectively. The changes taken in Xtend are described in
Ref.\citenum{2018SPIE10699E..23H, 2020SPIE11444E..23N, 2022SPIE12181E..1TM}.
Table~\ref{tab:Xtend-spec} summarizes specifications of Xtend. The SXI has four
CCDs, arranged in a 2$\times$2 grid, to achieve a large FoV of
$38^{\prime}\times38^{\prime}$. One CCD has a 1280$\times$1280 \textit{physical}
pixel format, which becomes a 640$\times$640 \textit{logical} pixel format through
on-chip 2$\times$2 binning. The logical pixel size is 48~$\mu$m~$\times$~48~$\mu$m,
which corresponds to $1.\!\!^{\prime\prime}77 \times 1.\!\!^{\prime\prime}77$. The
energy range is defined as 0.4--13~keV and well overlaps with that of Resolve. The
energy resolution is $\sim$180~eV (FWHM) at 5.9~keV, which is the average value of 4
CCDs and the individual values are shown in Section~\ref{sec:in-orbit}. The
effective area values in the table are derived from ground calibrations and will be
updated by ongoing in-orbit calibrations. The frame cycle is 4~seconds, during which
two halves of the imaging area, called segments, are separately and simultaneously
read out. We apply the charge injection (CI) technique to mitigate the charge loss due to
transfer inefficiency\cite{2009PASJ...61S...9U, 2014NIMPA.765..269N,
KANEMARU2020164646} and a CI row is placed every 80 logical rows.

\begin{center}       
 \begin{threeparttable}[ht]
  \caption{Observation modes of Xtend} \label{tab:mode}
  \begin{tabular}{|l|l|l|l|l|}
   \hline
   Mode & Fraction of read-out area & Exposure time & Number of exposure  & Live time \\
    & over imaging area & (sec)  & per frame & fraction$\dagger$ \\
   \hline
   Full window & 1 & 3.96 & 1 & 0.99 \\
   1/8 window & 1/8 & 0.46 & 8 & 0.93 \\
   1/8 window + burst & 1/8 & 0.06 & 8 & 0.12 \\
   Full window + burst & 1 & 0.06 & 1 & 0.015 \\
   \hline 
  \end{tabular}
\begin{tablenotes}
\item[$\dagger$] excluding charge transfer time, during which photons detected are recognized as trailing events
\end{tablenotes}
 \end{threeparttable}
\end{center}

Tab.~\ref{tab:mode} lists observation modes of Xtend. The full window mode is a
standard observation mode. The entire imaging area is read out per frame cycle. The
charge transfer time accounts for just about 1\% of the frame cycle and the live
time fraction (LTF) is about 99\%. The 1/8 window mode reads out only 1/8 of the
entire imaging area around on-axis position and repeats the exposure eight times per
frame cycle. Since the exposure time is reduced by a factor of 8, the pile-up
tolerance is increased by the same amount. The 1/8 window + burst mode reads out the
same area as the 1/8 window mode but reduces the exposure time down to 0.06 sec. The
pile-up tolerance is further increased in this mode in compensation for the loss of
LTF; the exposure time (0.06 sec) is much shorter than the read-out time for the 1/8
area (0.5 sec). The full window + burst mode is made for calibration purpose and not
open for guest observers. In this mode, the exposure time is significantly shorter
the read-out time for the entire imaging area so that the LTF is extremely low.

In terms of the observation mode, CCD1 and CCD2 are one pair, and CCD3 and CCD4 are
the other pair. Two CCDs in the same pair operate in the same observation mode, and
the two pairs can operate in different observation mode. The observation axis of
Xtend is located in CCD2, and the observation mode for the pair of CCD1 and CCD2 can
be selected by observers according to their observation purposes whereas that for
the pair of CCD3 and CCD4 is always the full window mode.

As is shown above, event pile-up is always a concern for CCD detectors. Therefore,
we developed the pile-up simulator dedicated for Xtend, which is presented in
Ref.~\citenum{YoneyamaSPIE2024}.

In our ground experiments, we developed a new observation mode, which is not listed
in Tab.~\ref{tab:mode}. The new observation mode is a countermeasure against an
issue we encountered in sub-system level tests in ground. We did not observe the
issue in the system-level tests in ground nor in orbit. We do not expect to use the
new observation mode in future for Xtend. The details of the issue and the new
observation mode are presented in Ref.~\citenum{NodaSPIE2024}.

\section{Performance verification in spacecraft thermal vacuum test}

\begin{figure} [htbp]
  \centering
  \includegraphics[width=0.7\textwidth]{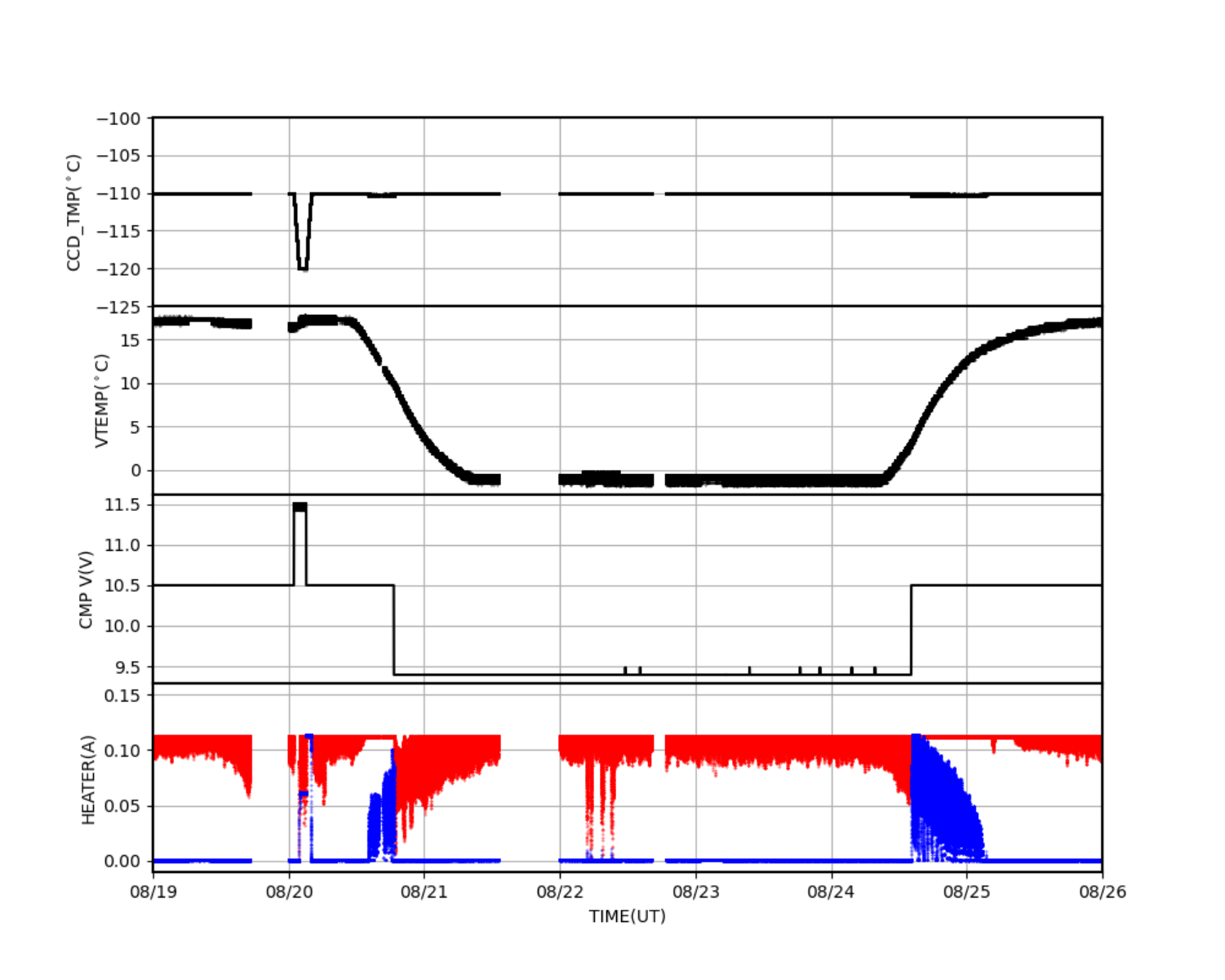}
  \caption[]{\label{fig:TV} Time histories of CCD temperature (top), video
  temperature (second from the top), cooler compressor voltage (third from the top),
  and heater currents (bottom).  Two heaters are attached to the plate on which CCDs
  are installed. Red and blue colors in the bottom plot indicates the two different
  heater currents.}
\end{figure}

After a series of performance verification tests in sub-system level, including the
thermal vacuum, vibration, and acoustic tests\cite{2022SPIE12181E..1TM}, the SXI
sub-system was installed to the XRISM spacecraft in April 2022. The spacecraft
system level tests, including initial electrical, thermal vacuum, mechanical
environmental, and final electrical tests, started in May 2022 and ended in March
2023\cite{TashiroSPIE2024}. Among these tests, the spacecraft thermal vacuum test
was the only occasion for us to cool the CCDs and verify the performances of
the SXI sub-system in detail. 

The spacecraft thermal vacuum test was performed over a period of a month in August
2022, during which the spacecraft experienced multiple thermal cycles.
Fig.~\ref{fig:TV} shows a time history of the CCD temperature for a given week along
with those of the video temperature, cooler compressor voltage, and heater
currents. Here, the video temperature is a kind of proxy of the thermal cycle phase;
in this period, the thermal cycle phase first shifted from the hot side to the cold
side and then shifted again from the cold side to the hot side. Through the swings
in the thermal environment, Xtend kept its CCD temperature constant at
$-110$~$^{\circ}$C, which is the initial operating temperature in orbit. The cooler
compressor voltage was adjusted so that the total duty cycle of the two heaters,
attached to the plate on which CCDs are installed, was around 50\%. It was also
verified that Xtend can cool the CCD down to $-120$~$^{\circ}$C in this period.

In this test, the spectroscopic performance was also verified by evaluating spectra
of X-ray events from the $^{55}$Fe calibration sources. In addition, adjustment of
the charge injection amount, confirmation of optical blocking power, and mock run of
nominal operation were also performed without major problems.

\section{Performances in orbit}
\label{sec:in-orbit}

Following the critical operation of the spacecraft, commissioning operation of the
bus components, and start-up operation of Resolve after the launch of XRISM, we
conducted the start-up operation of Xtend spending 5 days from October 17th, 2023.
This start-up operation went well as we tested on ground. All the subsequent
commissioning operations were also successfully performed. Here we summarize initial
results verifying the Xtend performance.

\begin{figure} [htbp]
  \centering \centering \includegraphics[width=0.5\textwidth]{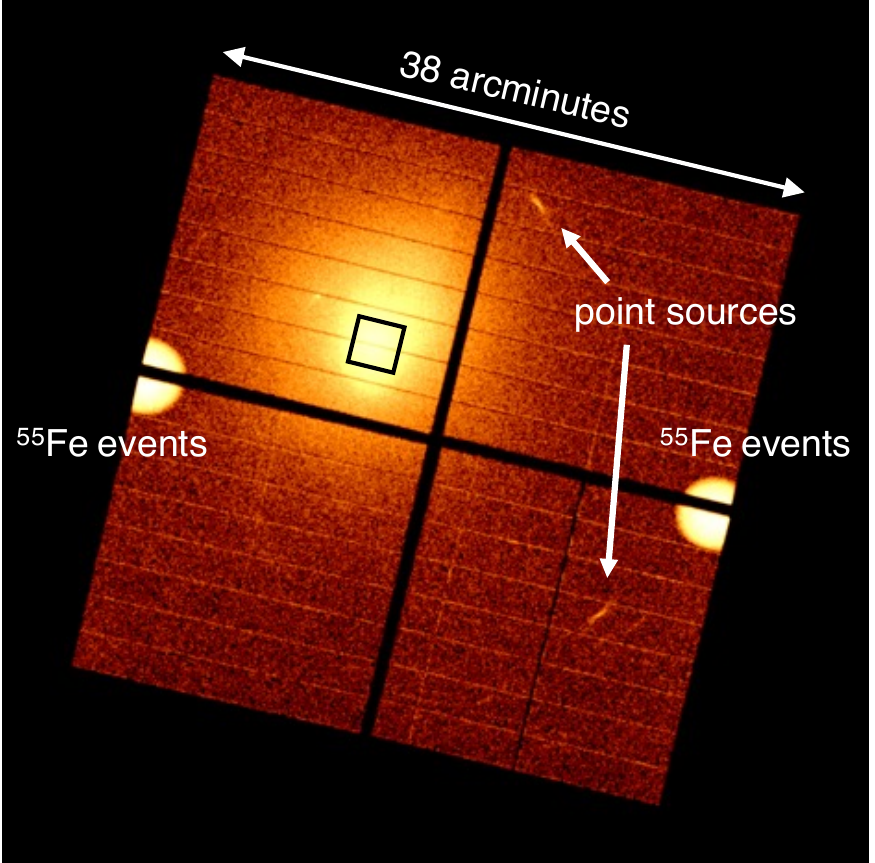}
  \caption[]{\label{fig:A2319} The first light image of Xtend. The galaxy cluster
  Abell~2319 and some off-axis point sources are captured. X-ray events from the
  $^{55}$Fe calibration sources are seen at the corners of CCDs. The black box
  indicates the FoV of Resolve.}
\end{figure}

Fig.~\ref{fig:A2319} shows the first light image of Xtend. The galaxy cluster
Abell~2319 is captured in this image. This is truly the fist light image in the
sense that XRISM was pointing to this galaxy cluster before and through the start-up
operation of Xtend. This image was made from the same data as the public first light
image released by JAXA, but without Vignetting correction, energy selection, or,
intentional color-scale adjustment so that this can be regarded as an unfiltered
image. A grade selection was applied. The entire region covering the
$38^{\prime}\times38^{\prime}$ FoV is fully functional for X-ray detection except
for bad columns known from ground tests, which is seen as a dark lane running at the
center of the CCD positioned at bottom right in this figure. Some off-axis point
sources are also detected. Their arc-like shape is due to the skewed PSF of the XMA
at off-axis.  The CCDs are not aligned perfectly parallel so that the width of the
gap between CCD active regions varies depending on the position from
$40^{\prime\prime}$ to $60^{\prime\prime}$. The corners of CCDs are irradiated with
X-ray from the $^{55}$Fe calibration sources. The count rates are $\sim$1~counts
sec$^{-1}$ CCD$^{-1}$ as of October 2023. The Xtend FoV completely encompasses the
Resolve FoV by shifting its geometrical center by $\sim$5$^{\prime}$. Please note
that the positions of the CI rows were later shifted by 35 rows not to interfere
with an on-axis source\cite{SuzukiSPIE2024}. The spacing of the CI rows was not
changed and remains at 80 rows.

\begin{figure} [htbp]
  \centering \centering
  \includegraphics[width=\textwidth]{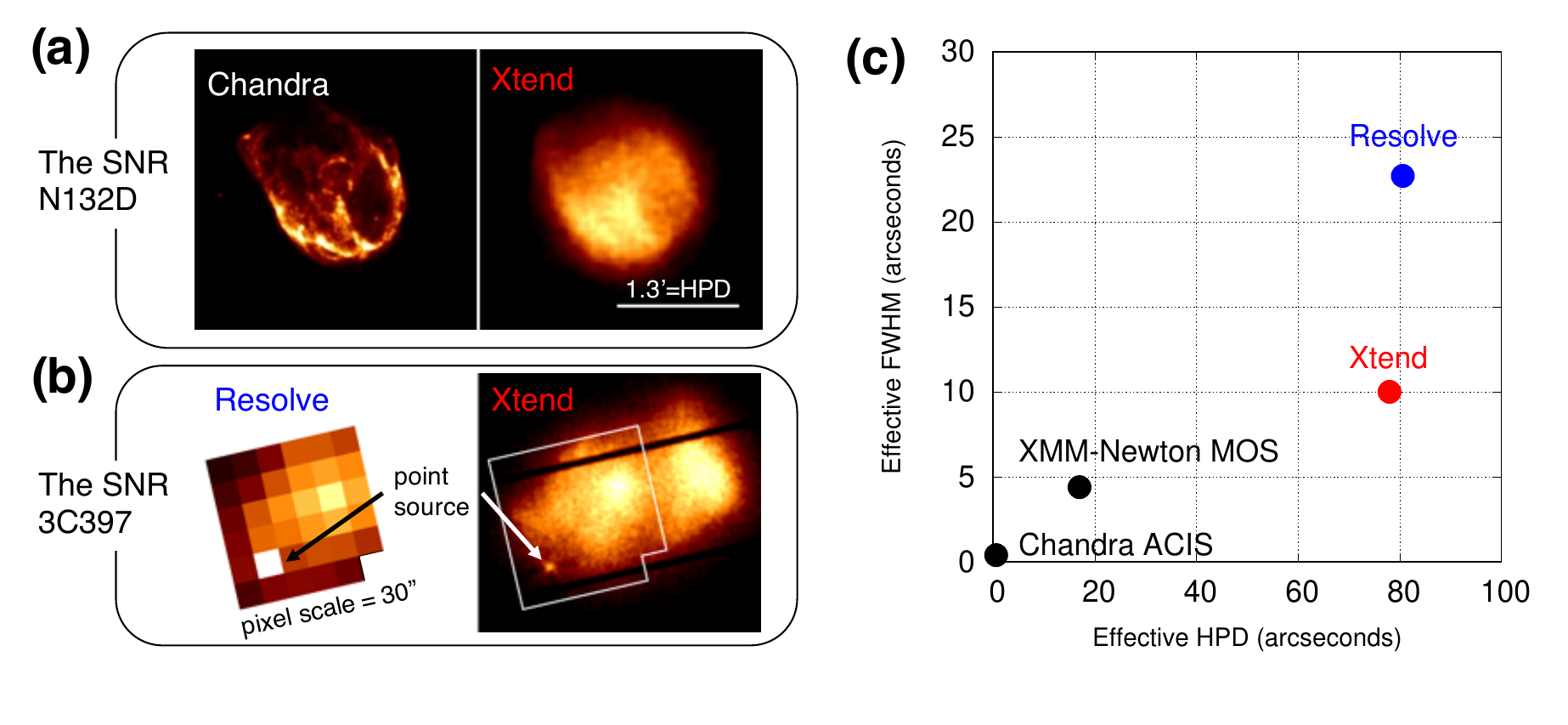}
  \caption[]{\label{fig:imagingPerformance} (a) Chandra and Xtend images of the
  supernova remnant N132D. The white bar indicates the size of $1.\!\!^{\prime}3$,
  which is the same as the HPD of the Xtend XMA\cite{TamuraSPIE2024}.  (b) Resolve
  and Xtend images of the supernova remnant 3C397. The Resolve FoV excluding the
  calibration pixel are superimposed on the Xtend image with a white polygon. A
  known background point source is detected in both images. (c) Relationships
  between the FWHM and HPD of point-source images obtained with Xtend (red) and
  Resolve (blue) as well as Chandra ACIS and XMM-Newton MOS (black).}
\end{figure}

Fig.~\ref{fig:imagingPerformance}(a) shows Chandra and Xtend images of the supernova
remnant N132D in the Large Magellanic Cloud. A three-color version of the Xtend
image was public-released by JAXA. In spite of that the diameter of this remnant is
almost the same as the HPD of the XMA of $1.\!\!^{\prime}3$, the internal
structures, like central bar and southwestern void, can be identified even in the
Xtend image. Fig.~\ref{fig:imagingPerformance}(b) shows Resolve and Xtend images of
the supernova remnant 3C397 in our Galaxy. A known background point source is
detected in both images. You can find that the position of the source can be more
precisely identified in the Xtend image. These characteristics can be understood
with Fig.~\ref{fig:imagingPerformance}(c), where relationships between the FWHM and
HPD of point-source images obtained with Xtend and Resolve as well as Chandra ACIS
and XMM-Newton MOS are plotted. Both the FWHM and HPD of a point-source image
reflect not only the X-ray mirror performance but also the blurring due to pixel
size of detectors, system attitude reconstruction inaccuracy, and so on. Although,
in terms of the effective HPD, the value of Xtend is about five times lager than
that of XMM-Newton, in terms of the effective FWHM, the value of Xtend is just twice
lager than that of XMM-Newton. The Xtend's sharp PSF, FWHM of
$\sim$10$^{\prime\prime}$, is beneficial to resolve spatial structures and detect
point sources. However, we need to keep it mind that the HPD is more important
measure to evaluate photon leakage from and to neighboring regions. In the
comparison between Resolve and Xtend, the difference in effective FWHM directly
reflects the difference in pixel scale because the FWHM of the XMA's PSF is in
between the pixel scales of Resolve and Xtend. In this regard, the effective HPDs of
Resolve and Xtend are more or less the same.

Regarding XMA-related items, the HPD and effective area of Xtend were also measured
to be consistent with results obtained in ground calibrations and satisfy mission
requirements\cite{TamuraSPIE2024}.

\begin{figure} [htbp]
  \centering \centering
  \includegraphics[width=\textwidth]{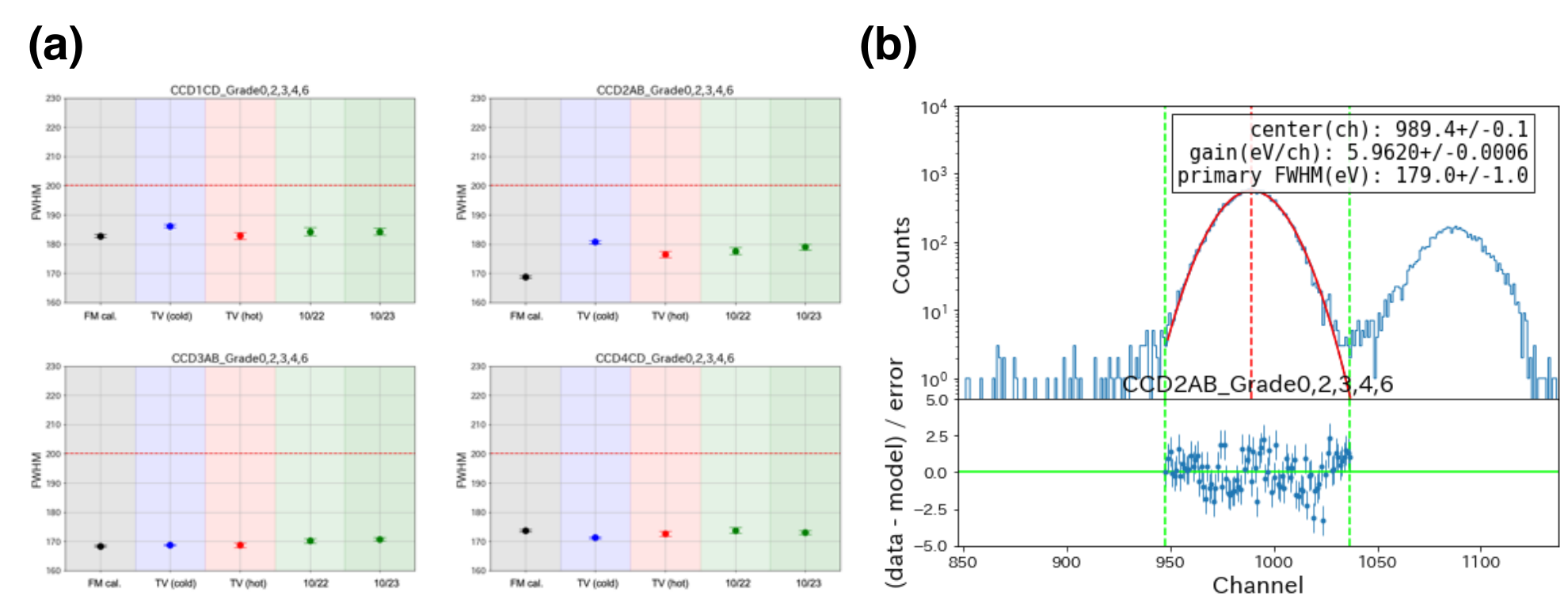}
 \vspace{0.1cm}
  \caption[]{\label{fig:spectroscopicPerformance} (a) Energy resolution measured in
  the calibration in the sub-system level test (black), the cold side (blue) and hot
  side (red) of spacecraft thermal vacuum test, and on October 22th (light green)
  and 23th (green), 2023 in orbit. Those of CCD1 (top left), CCD2 (top right), CCD3
  (bottom left), and CCD4 (bottom right) are shown. The red dotted line indicates
  the requirement imposed on Xtend at the beginning of life of XRISM. (b) Spectrum
  of X-ray events from the $^{55}$Fe calibration source taken with CCD2 (on-axis
  CCD). CTI corrections with parameters determined on ground are applied. The red
  curve is the best-fit single Gaussian to the Mn-K$\alpha$ line. }
\end{figure}

Fig.~\ref{fig:spectroscopicPerformance} summarizes energy resolution measured on
ground and in orbit. The Mn-K$\alpha$ line in the spectrum of X-ray events from the
$^{55}$Fe calibration sources is fitted with a single Gaussian and the FWHM derived
in the unit of eV are shown. CTI corrections with parameters determined on ground
are applied in the data analysis\cite{KANEMARU2020164646}. A single Gaussian
reasonably well fit the data with this level of statistics. The energy resolution
measured in orbit was $\sim$180~eV and satisfied the mission requirement of 200~eV
at the beginning of life. Results obtained in orbit are fully consistent with those
obtained on ground. In the case of the \Hitomi\ X-ray CCD camera, the energy
resolution measured in orbit was degraded by $\sim$10~eV in comparison with that
measured on ground\cite{2018PASJ...70...21N}. Although the reason of the degradation
was not clear, the \Hitomi\ spectrum might have been partly affected by a light leak
issue, which we took multiple measures in XRISM/Xtend\cite{2018SPIE10699E..23H,
UCHIDA2020164374}.

\begin{figure} [ht]
 \begin{minipage}{0.49\hsize}
  \centering \includegraphics[width=\textwidth]{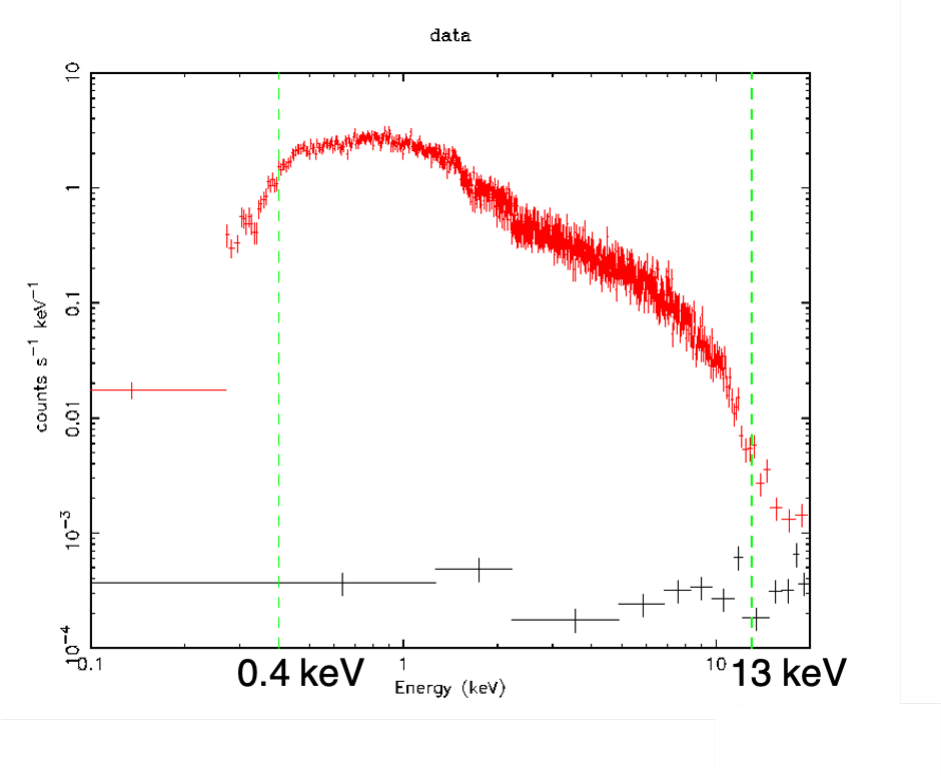}
  \caption[]{\label{fig:3C273} Xtend spectrum of the quasar 3C~273 (red) and its
  normalized NXB spectrum (black). The two vertical green dashed lines indicate
  the boundaries of Xtend's official energy range of 0.4--13~keV.}
 \end{minipage}
 \hspace{0.01\hsize}
  \begin{minipage}{0.49\hsize}
  \centering 
  \includegraphics[width=\textwidth]{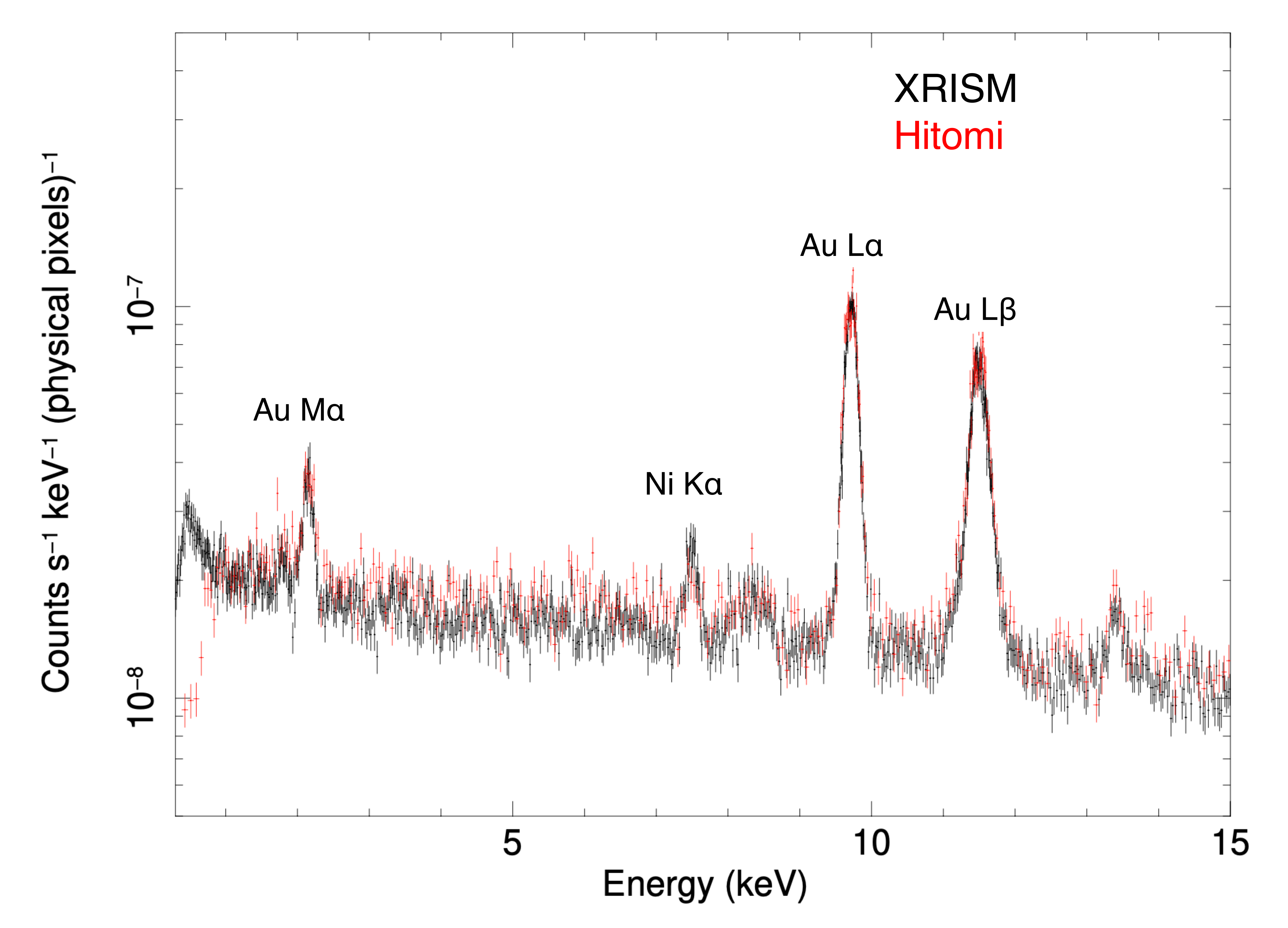}
  \caption[]{\label{fig:NXB} Non-X-ray background spectra of Xtend (black) and
  the \Hitomi\ X-ray CCD camera (red).}
 \end{minipage}
\end{figure}

Fig.~\ref{fig:3C273} shows the Xtend spectrum of the quasar 3C~273. This bright
quasar spectrum clearly demonstrates that Xtend covers the energy range of
0.4--13~keV. Actually, as you can see, Xtend is still sensitive to X-ray photons
beyond this energy band. We define this official energy band for which extensive
calibration work is performed. Calibration uncertainties in the linearity of energy
scale and the tail component of response function might be a concern above 13~keV
and below 0.4~keV, respectively.

Fig.~\ref{fig:NXB} shows the NXB spectra of Xtend and the \Hitomi\ X-ray CCD
camera. There is no major difference between the two spectra. This is as expected
because the orbit of the satellite and the design of the camera are the
same. Prominent lines are labeled in this figure. In addition, you can find an
apparent broad line at $\sim$8.3~keV and a weak line at $\sim$13.4 keV. The former
is likely a blend of Ni~K$\beta$ and Au~L$\iota$, and the latter is likely a
Au~L$\gamma$ blend. As stated above, in this spectrum, there is no strong lines
overlapping major lines from astronomical sources. This cleanness along with its low
and stable flux make it easy to model the NXB spectrum of Xtend and ensure its
reproducibility.

Lastly, we mention that there is no apparent contamination build-up as experienced
by the Suzaku CCD cameras\cite{suzakuTD} as of writing this paper, more than half a
year passed after the launch. We will continue to monitor this issue.

\section{SUMMARY}
\label{sec:summary}

Xtend is a soft X-ray imaging telescope developed for XRISM, consisting of of the
Soft X-ray Imager (SXI), an X-ray CCD camera, and the X-ray Mirror Assembly (XMA), a
thin-foil-nested conically approximated Wolter-I optics. Combining the SXI and XMA
with a focal length of 5.6~m, a field of view $38^{\prime}\times38^{\prime}$ of over
the energy range from 0.4 to 13 keV is realized. Xtend was successfully started up
40 days after the launch of the satellite and all the commissioning operations were
also successfully performed.  All the performances, including FoV,
HPD\cite{TamuraSPIE2024}, effective area\cite{TamuraSPIE2024}, energy resolution,
non-X-ray background, and contamination, are fully consistent with those measured in
ground tests and satisfy mission requirements. Now we are ready for science
outputs. Calibration activities are on-going.

\acknowledgments 

We acknowledge all the supports from Hamamatsu Photonics K.K., Mitsubishi Heavy
Industries Ltd., and Sumitomo Heavy Industries Ltd.\ to develop CCD, SXI system, and
cooler component, respectively, including those in the ASTRO-H era. This work was
supported by JSPS KAKENHI Grant Numbers 19K21884, 20H01947, 20KK0071, 23K20239,
21J00031, 22KJ3059, 24K17093, 21K20372, 23K22540, 22H01269, 21H04493, 24K17105,
21K03615, 24K00677, 21K13963, 23K22536, 21H01095, 23K20850, 23H00128, 20H00175,
20KK0071, 24H00253.


\bibliography{mybibfile} 

\begin{thebibliography}{10}

\bibitem{TashiroSPIE2024}
{Tashiro}, M. et~al., ``{Development and Operation Status of X-Ray Imaging and
  Spectroscopy Mission (XRISM)},'' {\em \it\procspie},  in press (2024).

\bibitem{2020SPIE11444E..22T}
{Tashiro}, M. et~al., ``{Status of x-ray imaging and spectroscopy mission
  (XRISM)},'' {\em \procspie} {\bf 11444},  1144422 (Dec. 2020).

\bibitem{2018JLTP..193..991I}
{Ishisaki}, Y. et~al., ``{Resolve Instrument on X-ray Astronomy Recovery
  Mission (XARM)},'' {\em Journal of Low Temperature Physics}~{\bf 193},
  991--995 (Dec. 2018).

\bibitem{2018SPIE10699E..23H}
{Hayashida}, K. et~al., ``{Soft x-ray imaging telescope (Xtend) onboard X-ray
  Astronomy Recovery Mission (XARM)},'' {\em \procspie} {\bf 10699},  1069923
  (July 2018).

\bibitem{HayashiSPIE2024}
{Hayashi}, T. et~al., ``{In-orbit performance of the XMA for XRISM/Resolve},''
  {\em \it\procspie},  in press (2024).

\bibitem{TamuraSPIE2024}
{Tamura}, K. et~al., ``{In-orbit performance of the Xtend-XMA onboard XRISM},''
  {\em \it\procspie},  in press (2024).

\bibitem{KellySPIE2024}
{Kelly}, R. et~al., ``{Overview and inflight performance of the resolve
  high-resolution soft X-Ray spectrometer on the X-Ray imaging and spectroscopy
  mission},'' {\em \it\procspie},  in press (2024).

\bibitem{PorterSPIE2024}
{Porter}, F. et~al., ``{In-flight performance of the XRISM/Resolve detector
  system},'' {\em \it\procspie},  in press (2024).

\bibitem{ozawa_2007}
{Ozawa}, H. et~al., ``{Development of p-type CCD for the NeXT: the next
  Japanese x-ray astronomical satellite mission},'' {\em \procspie} {\bf 6266},
   62662N (June 2006).

\bibitem{Takagi_2007}
{Takagi}, S. et~al., ``{Development of fully depleted and back-illuminated
  charge-coupled devices for soft x-ray imager onboard the NeXT satellite},''
  {\em \procspie} {\bf 6266},  62663V (June 2006).

\bibitem{Ueda_2011}
{Ueda}, S. et~al., ``{Development of the x-ray CCD for SXI on board ASTRO-H},''
  {\em \procspie} {\bf 8145},  814504 (Sept. 2011).

\bibitem{2018JATIS...4a1211T}
{Tanaka}, T. et~al., ``{Soft X-ray Imager aboard Hitomi (ASTRO-H)},'' {\em
  Journal of Astronomical Telescopes, Instruments, and Systems}~{\bf 4},
  011211 (Jan. 2018).

\bibitem{2022SPIE12181E..1TM}
{Mori}, K. et~al., ``{Xtend, the soft x-ray imaging telescope for the X-Ray
  Imaging and Spectroscopy Mission (XRISM)},'' {\em \procspie} {\bf 12181},
  121811T (Aug. 2022).

\bibitem{2018PASJ...70...21N}
{Nakajima}, H. et~al., ``{In-orbit performance of the soft X-ray imaging system
  aboard Hitomi (ASTRO-H)},'' {\em \pasj}~{\bf 70},  21 (Mar. 2018).

\bibitem{2017NIMPA.873...16N}
{Nakajima}, H. and {Hitomi Collaboration}, ``{Astronomical imaging with the
  X-ray observatory Hitomi},'' {\em Nuclear Instruments and Methods in Physics
  Research A}~{\bf 873},  16--20 (Nov. 2017).

\bibitem{TsuboiSPIE2024}
{Tsuboi}, Y. et~al., ``{X-ray transient search using XRISM/Xtend},'' {\em
  \it\procspie},  in press (2024).

\bibitem{Boissay-MalaquinSPIE2024}
{Boissay-Malaquin}, R. et~al., ``{X-ray Mirror Assembly for the X-Ray Imaging
  and Spectroscopy Mission (XRISM): comparison between ground calibration
  measurements and raytracing simulations},'' {\em \it\procspie},  in press
  (2024).

\bibitem{SuzukiSPIE2024}
{Suzuki}, H. et~al., ``{Initial operations of the Soft X-ray Imager onboard
  XRISM},'' {\em \it\procspie},  in press (2024).

\bibitem{2018JATIS...4a1213I}
{Iizuka}, R. et~al., ``{Ground-based x-ray calibration of the Astro-H/Hitomi
  soft x-ray telescopes},'' {\em Journal of Astronomical Telescopes,
  Instruments, and Systems}~{\bf 4},  011213 (Jan. 2018).

\bibitem{2020SPIE11444E..23N}
{Nakajima}, H. et~al., ``{Soft x-ray imager (SXI) for Xtend onboard X-Ray
  Imaging and Spectroscopy Mission (XRISM)},'' {\em \procspie} {\bf 11444},
  1144423 (Dec. 2020).

\bibitem{2009PASJ...61S...9U}
{Uchiyama}, H. et~al., ``{New CTI Correction Method for Spaced-Row Charge
  Injection of the Suzaku X-Ray Imaging Spectrometer},'' {\em \pasj}~{\bf 61},
  S9--S15 (Jan. 2009).

\bibitem{2014NIMPA.765..269N}
{Nobukawa}, K.~K. et~al., ``{Use of a charge-injection technique to improve
  performance of the Soft X-ray Imager aboard ASTRO-H},'' {\em Nuclear
  Instruments and Methods in Physics Research A}~{\bf 765},  269--274 (Nov.
  2014).

\bibitem{KANEMARU2020164646}
{Kanemaru}, Y. et~al., ``Experimental studies on the charge transfer
  inefficiency of ccd developed for the soft x-ray imaging telescope xtend
  aboard the xrism satellite,'' {\em Nuclear Instruments and Methods in Physics
  Research A}~{\bf 984},  164646 (2020).

\bibitem{YoneyamaSPIE2024}
{Yoneyama}, T. et~al., ``{Pile-up simulator for XRISM/Xtend},'' {\em
  \it\procspie},  in press (2024).

\bibitem{NodaSPIE2024}
{Noda}, H. et~al., ``{New CCD driving technique to suppress anomalous charge
  intrusion from outside the imaging area for soft X-ray imager of Xtend
  onboard XRISM},'' {\em \it\procspie},  in press (2024).

\bibitem{UCHIDA2020164374}
{Uchida}, H. et~al., ``Optical blocking performance of ccds developed for the
  x-ray astronomy satellite xrism,'' {\em Nuclear Instruments and Methods in
  Physics Research A}~{\bf 978},  164374 (2020).

\bibitem{suzakuTD}
{https://darts.isas.jaxa.jp/astro/suzaku/analysis/doc/suzaku\_td/}

\end{thebibliography}
\bibliographystyle{spiebib} 

\end{document}